\documentclass[galaxies,article,accept,moreauthors,10pt,a4paper]{Definitions/mdpi}
\graphicspath{{Definitions/}}

\usepackage{graphicx}
\usepackage{color}
\usepackage{url}
\usepackage{xspace}
\usepackage{amsmath}	% Advanced maths commands
\usepackage{amssymb}	% Extra maths symbols

\newcommand{\hen}{\mbox{Hen\,2-428}\xspace}
\newcounter{Rco}
\newcommand{\Ionst}[1]{\setcounter{Rco}{#1}\Roman{Rco}}
\newcommand{\Ion}[2]{\mbox{#1\,{\scriptsize\Ionst{#2}}}} % ie.\Ion{Fe}{2}
\newcommand{\Ionw}[3]{\mbox{#1\,{\scriptsize\Ionst{#2}}~$\lambda\,#3$\,\AA}} % ie. \Ionw{He}{2}{4686}
\newcommand{\Ionww}[3]{\mbox{#1\,{\scriptsize\Ionst{#2}}~$\lambda\lambda\,#3$\,\AA}} % ie. \Ionww{He}{2}{4200, 4542, 5412}
\newcommand{\logg}{\mbox{$\log g$}\xspace}
\newcommand{\loggw}[1]{\mbox{$\log g\hspace{-0.5mm} =\hspace{-0.5mm}  #1$}} % ie. \loggw{5.25 \pm 0.07}
\newcommand{\kK}{\mbox{\rm kK}\xspace}
\newcommand{\Teff}{\mbox{$T_\mathrm{eff}$}\xspace}
\newcommand{\Teffw}[1]{\mbox{$\Teff\hspace{-0.5mm} =\hspace{-0.5mm} #1 \,\mathrm{kK}$}} % ie. \Teffw{53.0 \pm 6.9}

\newcommand{\Msol}{$M_\odot$}

\firstpage{1}
\articlenumber{88}
\issuenum{3}
\pubvolume{6}
\pubyear{2018}
\copyrightyear{2018}
\history{Received: 19 June 2018; Accepted: 10 August 2018; Published: 14 August 2018}
\updates{yes}

\Title{Revealing the True Nature of Hen\,2-428}

\Author{Nicole Reindl $^{1,}$*$^{,\dagger}$, Nicolle L. Finch $^{1,\dagger}$, Veronika Schaffenroth $^{2}$, Martin A. Barstow $^{1}$, \linebreak Sarah L. Casewell $^{1}$, Stephan Geier $^{2}$, Marcelo M. Miller Bertolami $^{3}$ and Stefan Taubenberger $^{4}$}

\AuthorNames{Nicole Reindl, Nicolle L. Finch, Veronika Schaffenroth, Martin A. Barstow, Sarah L. Casewell, Stephan Geier, Marcelo M. Miller Bertolami and S. Taubenberger}

\address{%
$^{1}$ \quad Department of Physics and Astronomy, University of Leicester, University Road, Leicester LE1 7RH, UK;  nlf7@le.ac.uk (N.L.F.); martin.barstow@leicester.ac.uk (M.A.B.); slc25@leicester.ac.uk (S.L.C.)\\
$^{2}$ \quad Institute for Physics and Astronomy, University of Potsdam, Karl-Liebknecht-Str. 24/25, \linebreak 14476 Potsdam, Germany; schaffenroth@astro.physik.uni-potsdam.de (V.S.); sgeier@astro.physik.uni-potsdam.de (S.G.)\\
$^{3}$ \quad Instituto de Astrofísica de La Plata, UNLP-CONICET, La Plata, 1900 Buenos Aires, Argentina; mmiller@fcaglp.fcaglp.unlp.edu.ar\\
$^{4}$ \quad European Southern Observatory, Karl-Schwarzschild-Str. 2, D-85748 Garching, Germany; staubenb@partner.eso.org}

\corres{Correspondence: nr152@le.ac.uk; Tel.: +44-116-223-1385}

\firstnote{These authors contributed equally to this work.}
\abstract{\textls[-10]{The nucleus of \hen is a short orbital period (4.2\,h) spectroscopic binary, whose status as potential supernovae
type Ia progenitor has raised some controversy in the literature. We present preliminary results of a thorough analysis of
this interesting system, which combines quantitative non-local thermodynamic (non-LTE) equilibrium spectral modelling, radial velocity
analysis, multi-band light curve fitting, and state-of-the art stellar evolutionary calculations. Importantly, we~find that
the dynamical system mass that is derived by using all available {He} $\Roman{Rco}{\scriptsize \text{II}}$ lines does not exceed
the Chandrasekhar mass limit. Furthermore, the individual masses of the two central stars are too small to lead to an SN\,Ia in
case of a dynamical explosion during the merger process.}}

\keyword{binaries: spectroscopic; stars: atmospheres; stars: abundances; supernovae}

\begin{document}
%%%%%%%%%%%%%%%%%%%%%%%%%%%%%%%%%%%%%%%%%%
%% Only for the journal Gels: Please place the Experimental Section after the Conclusions

%%%%%%%%%%%%%%%%%%%%%%%%%%%%%%%%%%%%%%%%%%
%\setcounter{section}{-1} %% Remove this when starting to work on the template.
\section{Introduction}

The detection and study of progenitor systems of type Ia supernovae (SN Ia) are crucial to understand the exact
explosion mechanism of these important cosmic distance indicators. Although there is a general consensus that
only the thermonuclear explosion of a white dwarf can explain the observed features of those events, the nature
of their progenitor systems still remains elusive. In the single-degenerate model, a white dwarf accretes material
from a non-degenerate companion and explodes when it reaches the Chandrasekhar mass limit \cite{Whelan73}. An alternative
scenario is the double-degenerate model, in which the explosion is triggered during the merger process
of two white dwarfs \cite{Iben84, Webbink84, Shen2015}. Identifying progenitors for the latter scenario is particularly
interesting in view of the applicability of SN Ia as standardisable candles, as the merging system could exceed
or fall below the Chandrasekhar limit~significantly.

\citet{Santander15} have claimed to have discovered the first definite double-degenerate, super-Chandrasekhar system
that will merge within a Hubble time, namely the central stars of the planetary nebula (CSPN) \hen. They found a
photometric period of 4.2\,h and that \Ionw{He}{2}{5411} is double lined and time variable.
By fitting the radial velocities (RVs) and light curves, they~concluded that the system consists of two pre-white dwarfs
with equal masses of 0.88\,\Msol.~In this case, the~system would merge within 700 million years making, \hen one of
the best SN Ia progenitor candidates~known.

This scenario has since been challenged by \citet{Garcia-Berro16}, who criticized the strong mismatch between the
luminosities and radii of both pre-white dwarf components as derived by \cite{Santander15} with the predictions
from stellar evolution models \cite{Bloecker91}. In addition, Reference \cite{Garcia-Berro16} suggested that the variable
\Ionw{He}{2}{5411} line might instead be a superposition of an absorption line plus an emission line, possibly
arising from the nebula, the irradiated photosphere of a close companion, or a stellar wind. Since this would question
the dynamical masses derived by \cite{Santander15}, Reference \cite{Garcia-Berro16} that repeated the light curve fitting and showed
that the light curves of \hen may also be fitted well by assuming an over-contact binary system that consists of two
lower mass (i.e., masses of 0.47\,\Msol\ and 0.48\,\Msol) stars. Thus, Reference \cite{Garcia-Berro16} concludes that the claim
that \hen provides observational evidence for the double degenerate scenario for SN Ia is~premature.

Given the potential importance of \hen as a unique laboratory to study the double degenerate merger scenario, it is highly
desirable to resolve this debate. Therefore, we use an improved approach for the analysis of this unique object by
combining quantitative non-LTE spectral modelling, RV~analysis, multi-band light curve fitting, and state-of-the art
stellar evolutionary calculations.

%%%%%%%%%%%%%%%%%%%%%%%%%%%%%%%%%%%%%%%%%%

\section{Spectral Analysis}

\begin{figure}[H]
\centering
\includegraphics[scale=0.65]{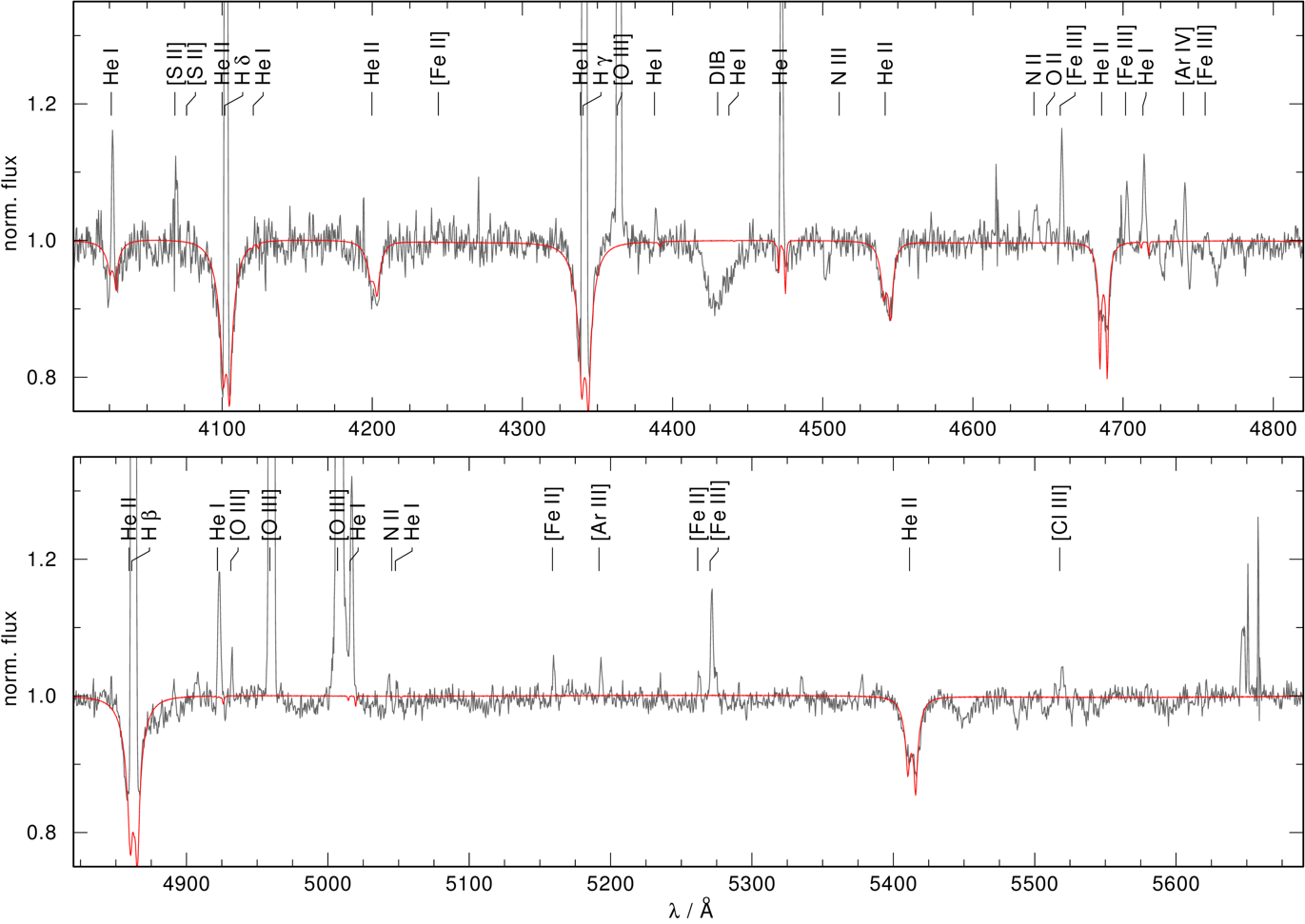}
\caption{\label{fig:fit} The best fit TMAP model (\textbf{red}) plotted against the OSIRIS spectrum (\textbf{grey}). Identified photospheric
%Define if appropriate
absorption lines, a diffuse interstellar band, and nebular emission lines are marked.}
\end{figure}

Our spectral analysis is based on the Very Large Telescope/FOcal Reducer and low dispersion Spectrograph 2 (VLT/FORS2)
and Gran Telescopio Canarias/Optical System for Imaging and low-Intermediate-Resolution Integrated Spectroscopy
(GTC/OSIRIS) spectra. FORS2 spectra were downloaded from the European Southern Observatory (ESO) archive
(ProgIDs 085.D-0629(A), 089.D-0453(A)) and by reduced by using standard IRAF procedures.~Calibrated GTC/OSIRIS spectra
were obtained from the GTC PublicArchive (ProgID GTC41-13A). Since the relatively broad and deep absorption lines in the
spectra of \hen cannot be fitted assuming a single component, we conclude that \hen is indeed a double lined spectroscopic
binary. We employed the T\"{u}bingen Model Atmospheres Package (TMAP, \cite{Werner03,TMAP,Rauch03}) to produce state-of-the-art
non-LTE model atmospheres. Initially, our models contain only H and He. The~model grid spans from \Teff\,$=$\,30.0--70.0\,\kK
(2.5\,\kK~steps), \loggw{4.25}--6.00 (0.25 steps), and covers three He abundances ($\log\,$(He/H) $ = -2, -1, 0$, number
fractions).
The first results of this analysis based on the FORS2 spectra only were presented in \cite{Finch18}. We~extended this
analysis by fitting the OSIRIS spectra and considering a variable flux contribution of the two stars in our
fitting procedure. To~constrain the parameters of the system, we used the XSPEC software \cite{Shafer91, xspec}, a
$\chi^2$ minimisation code, originally designed for X-ray spectra, but which has been adapted to work on optical
data \cite{Dobbie04}. XSPEC determines the best fit model for the input parameters, which in our case are the
effective temperatures (\Teff), surface gravities ($\log g[\mathrm{cm/s^2}]$), He abundances, RVs, and the flux
contribution of each star. First, the radial velocity of each component was found, and then these were fixed whilst
deriving \Teff, \logg, $\log$\,(He/H), and the flux contribution of each star simultaneously.
For the first star, we find \Teffw{48 \pm 7}, \loggw{5.00 \pm 0.1}, $\log\,$(He/H)$ = -1.1$ and relative flux contribution
of 46\%. For the second star, we derive \Teffw{46 \pm 7}, \loggw{4.8 \pm 0.1}, $\log\,$(He/H)$ = -1.0$ and relative flux
contribution of 54\%. The reduced $\chi^2$ value for these parameter is 1.04. In Figure~\ref{fig:fit}, we show the best
fit model (red) compared to an OSIRIS spectrum (grey).
We~obtain a good fit for \Ionww{He}{2}{4200, 4542, 5412}, and also the absorption wings of the \Ion{He}{1}
and \Ion{H}{1} lines are reproduced nicely.~The line cores of \Ionw{He}{2}{4686}, however, appear too deep compared to
the observation. This is a known problem when fitting the spectra of CSPNe and other hot stars with pure
HHe models only. It has been shown that \Ionw{He}{2}{4686} is particularly susceptible to metal line blanketing
\cite{Reindl14, Latour15}. We expect the systematic effects introduced by neglecting metal-line blanketing effects
to be of the order of 0.1\,dex on \logg and 1\,kK on \Teff \cite{Latour2010}, but we will include metals in our models for
future analysis of this system to overcome this problem.

\section{Light Curve Modelling}

\begin{figure}[H]
\centering
\includegraphics[scale=0.43]{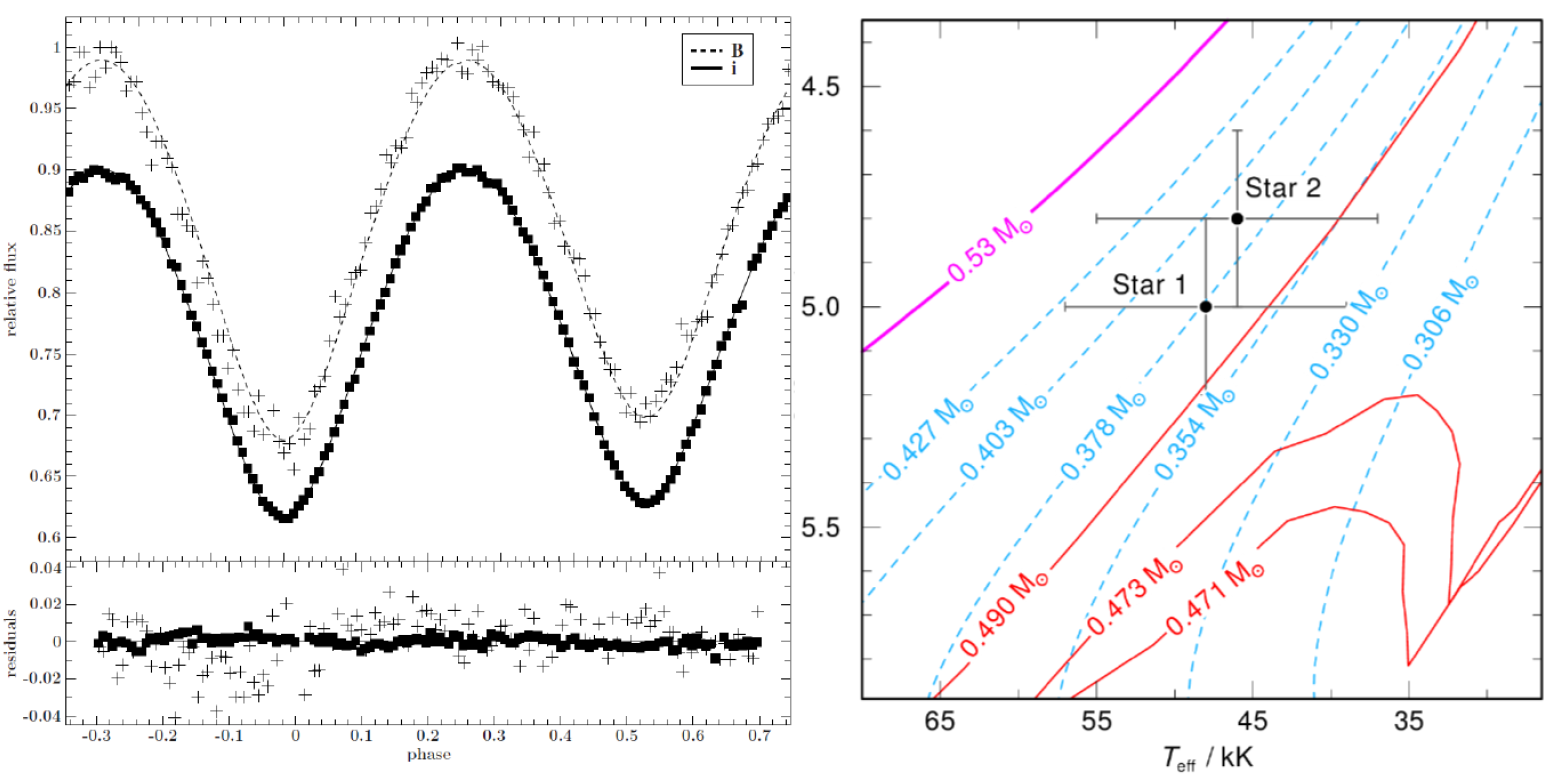}
\caption{\label{fig:evo} (\textbf{left}): Johnson B-band (dotted line) and Sloan i-band (solid line) light curves and LCURVE models; (\textbf{right}): the location of the two CSPNe of Hen 2-428 (black dots) in the \Teff\--\logg\ plane compared with a H-rich post-AGB evolutionary track
(pink, \cite{Bertolami16}), post-RGB tracks (dashed blue, \cite{Hall13}), and post-EHB tracks (red, \cite{Dorman1993}). The tracks
are labelled with stellar masses.}
%Define if appropriate
\end{figure}

The analysis of the light curves (downloaded from \cite{Santander15}) was carried out simultaneously in B and i filter.
For the analysis, we used MORO (Modified Roche Program, see \cite{Drechsel1995}), which is based on the
Wilson--Devinney code but is using a modified Roche model considering the influence of the radiation pressure on the
shape of the stars. Due to the significant degeneracies in the parameters found in the analysis of the light curve,
which result from not many independent parameters being used in the light curve analysis, we fixed the mass ratio of the
system to the mass ratio, which was derived by the analysis of the radial velocity curve. Moreover, we used the
temperatures derived by the spectral analysis as starting values. Initial attempts to reproduce the light curves
with a contact system failed. The~shape of the light curve could not be reproduced. Therefore, we tried to fit an
over-contact system, which~is assuming equal Roche potentials for both stars. We also considered a third light
source since Hen\,2-428 is known to exhibit a red-excess, which possibly results from the PN and a~distant companion~\cite{Rodriguez01}.
By varying the inclination, temperatures, Roche potentials, and luminosity ratio of both stars, the
light curves could be reproduced nicely. Our preliminary best fit to the light curves is shown in Figure~\ref{fig:evo}.
We get a relative luminosity for the primary of 53.8\% in B and 53.5\% in i. We derived an additional constant flux
component of 4.3\% in B and 10.3\% in i. We note that our fit reproduces the light curves better than the one of
\cite{Garcia-Berro16}, and also slightly better than the model of \cite{Santander15}.

\section{Dynamical Masses}

The amplitudes of the RV curves ($K_1$, $K_2$) were obtained by sinusoidal fitting of the
individual RV measurements obtained from the OSIRIS spectra of both components of the binary. This was done separately
in fine steps over a range of test periods within the uncertainties of the orbital period ($0.1758\pm0.0005$\,d) derived
from photometry by \cite{Santander15}. For each period, the $\chi^2$ of the best fitting sine curve was determined and the
solution with the lowest $\chi^2$ was chosen.
We can reproduce the results of \cite{Santander15} using their RVs as derived by merely Gaussian fitting of the
\Ionw{He}{2}{5412} absorption lines (fifth column in Table~\ref{tab:masses}). Somewhat lower RV amplitudes are obtained
when using our RVs as measured with XSPEC from our synthetic spectra for \Ionw{He}{2}{5412} (forth column). Surprisingly,
we~find lower RV amplitudes when using only the RVs from \Ionw{He}{2}{4686} (third column), and when all four \Ion{He}{2}
lines were fitted simultaneously (first and second column).
Since the inclination of the system ($i=65.3\pm0.1$) can be constrained well from the light curve fitting, the dynamical
masses of the two stars (bottom two rows in Table~\ref{tab:masses}) can be calculated via the binary mass function.
Importantly, we find that the combined dynamical mass obtained using all four \Ion{He}{2} lines is considerably lower
than the one reported by \cite{Santander15} and does not exceed the Chandrasekhar mass limit.

\begin{table}[H]
\caption{Dynamical and spectroscopic masses as derived in this work and by SG + 2015 \cite{Santander15}.}
%Define if appropriate
\centering
%% \tablesize{} %% You can specify the fontsize here, e.g.  \tablesize{\footnotesize}. If commented out \small will be used.
\begin{tabular}{cccccc}
\toprule
& \textbf{All Lines}	& \pmb{\Ionw{He}{2}{4686}} & \pmb{\Ionw{He}{2}{5412}} & \pmb{\Ionw{He}{2}{5412}} & \textbf{SG + 2015}\\
& \textbf{Syn. Spectra}	& \textbf{Syn. Spectra} & \textbf{Syn. Spectra} & \textbf{Gaussians} & \textbf{Gaussians} \\
\midrule
$K_1$ [km/s] & 174 $\pm$ 16 & 171 $\pm$ 18 & 196 $\pm$ 19 & 203 $\pm$ 13 & 206 $\pm$ 8 \\
$K_2$ [km/s] & 174 $\pm$  8 & 170 $\pm$ 10 & 191 $\pm$ 13 & 202 $\pm$ 12 & 206 $\pm$ 12 \\
\midrule
M$_1$ [\Msol] & 0.51 $\pm$ 0.15 & 0.48 $\pm$ 0.16 & 0.70 $\pm$ 0.23 & 0.81 $\pm$ 0.25 & 0.88 $\pm$ 0.13 \\
M$_2$ [\Msol] & 0.51 $\pm$ 0.14 & 0.48 $\pm$ 0.15 & 0.71 $\pm$ 0.22 & 0.81 $\pm$ 0.24 & 0.88 $\pm$ 0.13 \\
\bottomrule
\end{tabular}
\label{tab:masses}
\end{table}

%%%%%%%%%%%%%%%%%%%%%%%%%%%%%%%%%%%%%%%%%%
\section{Discussion}

Our spectral analysis is hindered by the limited number of unblended photospheric lines and the occurrences of metal line
blanketing effects. Higher S/N spectra that cover other unblended photospheric lines as well as models that also
include metal opacities would allow us to reduce the errors on the atmospheric parameters of the stars. Our effective
temperatures are significantly larger than the ones reported by \cite{Santander15}, who derived the stellar temperatures
from light curve fitting. The~narrow temperature range (30--40\,kK) adopted by \cite{Santander15}, however, is not valid
as already pointed out by \cite{Garcia-Berro16}. Ref.~\cite{Santander15} established the upper limit of 40\,kK based on the absence
\Ion{He}{2} emission lines, but~there are many CSPNe with even higher \Teff\ and also a lacking \Ion{He}{2} nebular lines.
Our light curve analysis shows that the light curves can be fitted equally well with higher \Teff values compared to the ones
derived by~\cite{Santander15}. This suggests that, because of the light curve solution degeneracy, light curve modelling alone
is not sufficient to derive the temperatures of the stars. Furthermore, our light curve modelling supports the idea that
\hen is an over-contact system, i.e., that both stars still share a common envelope. Therefore, should the PN be the result
of the common envelope ejection, future studies of this system could reveal important insights on the common envelope ejection
efficiency.

\textls[-15]{The dynamical masses are the main key to revealing the nature of \hen, i.e., to find out about the nature of the progenitor stars
as well as its SNIa status. We found that the system masses vary depending on which \Ion{He}{2} lines were used, but agree
within the error limits. Importantly, we find that the combined dynamical mass obtained using all four \Ion{He}{2} lines
does not exceed the Chandrasekhar mass limit and that the individual masses of the two CSPNe are too small for a reasonable
production of $^{56}$Ni (which determines the explosion brightness) in case of a dynamical explosion during the merger process.
Thus,~the~merging event of \hen would not be identified as a SN\,Ia (see Figure~3 in \cite{Shen2015}). }

The reason why the analysis of the \Ionw{He}{2}{5412} leads to a higher measurement of the RV amplitudes is unclear at the
moment. Reference~\cite{Rodriguez01} reported an excess that shows up red-wards of 5000\,\AA, possibly originating
from a late type companion. We speculate that this red-excess might impact the \Ionw{He}{2}{5412}, but not the other three \Ion{He}{2}
lines detected. Spectra extending toward longer wavelengths would help investigate the red excess further.

The lower system mass solution is also supported in view of the evolutionary time scales. The~low surface gravities of both
stars indicate that neither of the two stars has entered the white dwarf cooling sequence yet. The heating rate of post-asymptotic
giant branch (AGB) stars is strongly mass dependent. For post-AGB stars with masses greater than about 0.7\,\Msol, the
blue-ward evolution in the HRD is predicted to be so rapid that changes of \Teff\, would become noticeable within a
human life span. Recent evolutionary models \cite{Bertolami16} predict that, for a 0.71\,\Msol\ post-AGB star, it takes only
20 years to heat up from 30\,kK (that is when the nebula becomes visible) to 50\,kK (approximately the \Teff\ of both stars
now). Since \hen has been known for more than 50 years \cite{Hen1967}, this provides further evidence that the two CSPNe
cannot both be such massive post-AGB stars. We also note that, if the two stars were in thermal non-equilibrium after the
common-envelope was ejected, an even faster evolution would be~predicted.

The spectroscopic masses of the two stars (Figure~\ref{fig:evo}) agree with the dynamical masses from all four \Ion{He}{2} lines.
We stress, however, that the spectroscopic masses should be treated with caution because it is not clear to what extent the
evolutionary tracks are altered for over-contact systems. We~also note that radii from the best light curve fit are
significantly larger than the radii derived by the spectroscopic analysis due to the over-contact nature of the system. The
errors on both the dynamical and spectroscopic masses of the two stars are relatively large. While the spectroscopic masses
lead to the speculation that the two CSPNe might be AGB-manqu\'{e} stars (stars that fail to evolve through the AGB), the
dynamical masses do not allow us to distinguish whether one (or both) stars are post-AGB or post-red giant branch (RGB) stars.
Potentially, one of them could be a post-extreme horizontal branch (EHB) star. More precise RVs and a more accurate orbital
period would, therefore, be desirable to put stronger constraints on the masses.

\section{Conclusions}

Our preliminary results suggest that the dynamical system mass of Hen\,2-428 that is derived by using all available
\Ion{He}{2} lines does not exceed the Chandrasekhar mass limit. Furthermore, the
individual masses of the two central stars are also too small to lead to an SN\,Ia in case of a dynamical explosion
during the merger process. Further investigations on the red-excess and atmospheric models that consider opacities
of heavy metals are mandatory for a reliable analysis this intriguing system.

%%%%%%%%%%%%%%%%%%%%%%%%%%%%%%%%%%%%%%%%%%
\vspace{6pt}

\authorcontributions{Writing---Original Draft Preparation, N.R.; Writing---Review and Editing, N.R., N.L.F, V.S., M.A.B., S.L.C., S.G., M.M.M.B., and S.T.; Visualization, N.R. and V.S.; Supervision, N.R., S.L.C., and M.A.B.; Data reduction: S.T.; Spectral Analysis: N.L.F. and N.R.; Light curve modeling: V.S.; RV analysis: N.L.F. and S.G.; mass determination: N.R.; Interpretation of the results: N.R., S.G., V.S., M.M.M.B, S.T.}

\funding{N.R. is supported by a RC1851 research fellowship. N.L.F. is supported by an Science and Technology Facilities Council Studentship.}

%%%%%%%%%%%%%%%%%%%%%%%%%%%%%%%%%%%%%%%%%%
\acknowledgments{This work is based on observations collected at the European Organisation for Astronomical
Research in the Southern Hemisphere under ESO programmes 085.D-0629(A) and 089.D-0453(A) based on data from
the GTC PublicArchive at CAB (INTA-CSIC). The T\"ubingen Model-Atom Database (TMAD) service
(\url{http://astro-uni-tuebingen.de/~TMAD}) used to compile atomic data for this work was constructed as part
of the activities of the German Astrophysical Virtual Observatory. This research used the SPECTRE and ALICE
High Performance Computing Facilities at the University of Leicester.}

\conflictsofinterest{The authors declare no conflict of interest. The founding sponsors had no role in the design of the study; in the collection, analyses, or interpretation of data; in the writing of the manuscript, and in the decision to publish the results.}

%%%%%%%%%%%%%%%%%%%%%%%%%%%%%%%%%%%%%%%%%%
% Citations and References in Supplementary files are permitted provided that they also appear in the reference list here.

%=====================================
% References, variant A: internal bibliography
%=====================================
\reftitle{References}

% The following MDPI journals use author-date citation: Arts, Econometrics, Economies, Genealogy, Humanities, IJFS, JRFM, Laws, Religions, Risks, Social Sciences. For those journals, please follow the formatting guidelines on http://www.mdpi.com/authors/references
% To cite two works by the same author: \citeauthor{ref-journal-1a} (\citeyear{ref-journal-1a}, \citeyear{ref-journal-1b}). This produces: Whittaker (1967, 1975)
% To cite two works by the same author with specific pages: \citeauthor{ref-journal-3a} (\citeyear{ref-journal-3a}, p. 328; \citeyear{ref-journal-3b}, p.475). This produces: Wong (1999, p. 328; 2000, p. 475)

%%%%%%%%%%%%%%%%%%%%%%%%%%%%%%%%%%%%%%%%%%

\begin{thebibliography}{999}
%% Reference 1
%\bibitem[Author1(year)]{ref-journal}
%Author1, T. The title of the cited article. {\em Journal Abbreviation} {\bf 2008}, {\em 10}, 142-149, DOI.
%% Reference 2
%\bibitem[Author2(year)]{ref-book}
%Author2, L. The title of the cited contribution. In {\em The Book Title}; Editor1, F., Editor2, A., Eds.; Publishing House: City, Country, 2007; pp. 32-58, ISBN.

% Authors does not write the titles of the cited reference. Added by Tennyson Pei. 2018.8.10
%1
\bibitem[Whelan and Iben, 1973]{Whelan73} Whelan, J.; Iben, I., Jr. Binaries and Supernovae of Type I. \textit{Astrophys. J.} \textbf{1973}, \emph{186}, 1007--1014. [\href{http://dx.doi.org/10.1086/152565}{CrossRef}]

%2
\bibitem[Iben and Tutukov, 1984]{Iben84} Iben, I., Jr.;  Tutukov, A.V. Supernovae of type I as end products of the evolution of binaries with components of moderate initial mass (M not greater than about 9 solar masses). \emph{Astrophys. J. Suppl. Ser.} \textbf{1984}, \emph{ 54}, 335--372. [\href{http://dx.doi.org/10.1086/190932}{CrossRef}]

%3
\bibitem{Shen2015} 
\textls[-10]{Shen, K.J. Every Interacting Double White Dwarf Binary May Merge. \emph{ Astrophys. J. Lett. }\textbf{2015},  {\emph{805}, L6}. [\href{http://dx.doi.org/10.1088/2041-8205/805/1/L6}{CrossRef}]}
%4
\bibitem[Webbink, 1984]{Webbink84} Webbink, R.F. Double white dwarfs as progenitors of R Coronae Borealis stars and Type I supernovae. \textit{Astrophys. J.} \textbf{1984},  \emph{277}, 355--360. [\href{http://dx.doi.org/10.1086/161701}{CrossRef}]

%5
\bibitem[Santander-Garcia et al., 2015]{Santander15} 
\textls[-20]{Santander-Garcia, M.; Rodriguez-Gil, P.; Corradi, R.L.M.; Jones, D.; Miszalski, B.; Boffin, H.M.J.;  Rubio-Díez,~M.M.;   Kotze, M.M. The double-degenerate, super-Chandrasekhar nucleus of the planetary nebula Henize 2-428. \textit{Nature} \textbf{2015},  \emph{519}, 63--65.} [\href{http://dx.doi.org/10.1038/nature14124}{CrossRef}] [\href{http://www.ncbi.nlm.nih.gov/pubmed/25686608}{PubMed}]
%6
\bibitem[Garcia-Berro et al., 2016]{Garcia-Berro16} Garcia-Berro, E.; Soker, N.; Althaus, L.G.; Ribas, I.;  Morales, J.C. Is the central binary system of the planetary nebula Henize 2--C428 a type Ia supernova progenitor?  \textit{New Astron.}  \textbf{2016}, \emph{45}, 7--13. [\href{http://dx.doi.org/10.1016/j.newast.2015.10.010}{CrossRef}]

%7
\bibitem[Bl\"{o}cker and Sch\"{o}nberner, 1991]{Bloecker91} 
\textls[-10]{Bl\"{o}cker, T.; Sch\"{o}nberner, D. A 7-solar-mass AGB model sequence not complying with the core mass-luminosity relation. \textit{Astron. Astrophys.} \textbf{1991},  \emph{244}, L43--L46.}

%8
\bibitem[Rauch and Deetjen, 2003]{Rauch03} Rauch, T.; Deetjen, J.L. Handling of Atomic Data.~In  Proceedings of the 19th European Workshop on White Dwarfs, T\"ubingen,~Germany, 8--12 April 2002.
%9
\bibitem[Werner et al., 2003]{Werner03} Werner, K.; Deetjen, J.L.; Dreizler, S.; Nagel, T.; Rauch, T.; Schuh, S.L. Model Photospheres with Accelerated Lambda Iteration. In Proceedings of the 19th European Workshop on White Dwarfs, T\"ubingen, Germany, 8--12 April 2002; pp. 31--50.

%10
\bibitem[Werner et al., 2012]{TMAP} Werner, K.; Dreizler, S.; Rauch, T. \emph{TMAP: Tübingen NLTE Model-Atmosphere Package}; University of Tübingen: Tübingen, Germany, {2012}.%Please add city and country.


%11
\bibitem[Finch et al., 2018]{Finch18} Finch, N.L.; Reindl, N.; Barstow, M.A.; Casewell, S.L.; Geier, S.; Bertolami, M.M.M.;  Taubenberger, S. Spectral analysis of the binary nucleus of the planetary nebula Hen 2-428--first results. \textit{Open Astron.} \textbf{2018},  \emph{27}, 57--61. [\href{http://dx.doi.org/10.1515/astro-2018-0017}{CrossRef}]


%12
\bibitem[Arnaud et al., 2017]{xspec} Arnaud, K.; Gordon, C.;  Dorman, B. \textit{An X-ray Spectral Fitting Package: User's Guide for Version 12.9.1}; National Aeronautics and Space Administration: Greenbelt, MD, USA, {2017}.
%13
\bibitem[Shafer et al., 1991]{Shafer91} Shafer, R.A.; Haberl, F.; Arnaud, K.A. \textit{XSPEC: An X-Ray Spectral Fitting Package: Version 2 of the User's Guide}; National Aeronautics and Space Administration: Greenbelt, MD, USA, {1991}.

%14
\bibitem[Dobbie et al., 2004]{Dobbie04} Dobbie, P.D.; Pinfield, D.J.; Napiwotzki, R.; Hambly, N.C.; Burleigh, M.R.; Barstow, M.A.;  Jameson, R.F.; Hubeny, I.  Praesepe and the seven white dwarfs. \emph{Mon. Not. R. Astron. Soc.}\textbf{2004}, \textit{355}, L39--L43. [\href{http://dx.doi.org/10.1111/j.1365-2966.2004.08522.x}{CrossRef}]
%15
\bibitem[Latour et al., 2015]{Latour15} Latour, M.; Fontaine, G.; Green, E.M.;  Brassard, P. A non-LTE analysis of the hot subdwarf O star \mbox{BD + 28--4211-II}. The optical spectrum. \textit{Astron. Astrophys.} \textbf{2015},  \emph{579}, A39. [\href{http://dx.doi.org/10.1051/0004-6361/201525999}{CrossRef}]

%16
\bibitem[Reindl et al., 2014]{Reindl14} Reindl, N.; Rauch, T.; Werner, K.; Kruk, J.W.; Todt, H. On helium-dominated stellar evolution: The mysterious role of the O(He)-type stars. \textit{Astron. Astrophys.} \textbf{2014}, \emph{566}, A116. [\href{http://dx.doi.org/10.1051/0004-6361/201423498}{CrossRef}]


%17
\bibitem[Latour et al., 2010]{Latour2010} Latour M., Fontaine G., Brassard P.; Chayer, P.; Green, E.M.  A NLTE model atmosphere analysis of the pulsating sdO star SDSS J1600+0748. \textit{Astrophys. Space Sci.}  \textbf{2010}, \emph{329}, 141--144. [\href{http://dx.doi.org/10.1007/s10509-010-0343-9}{CrossRef}]

%18

\bibitem{Drechsel1995} Drechsel,  H.; Haas,  S.; Lorenz,  R.; Gayler, S. Radiation pressure effects in early-type close binaries and implications for the solution of eclipse light curves. \textit{Astron. Astrophys.} \textbf{1995},  \emph{294}, 723--743.
%19
\bibitem[Rodriguez et al., 2001]{Rodriguez01} Rodriguez, M.; Corradi, R.L.M.; Mampaso, A. Evidence for binarity in the bipolar planetary nebulae A 79, He 2-428 and M 1-91. \textit{Astron. Astrophys.} \textbf{2001},  \emph{377}, 1042--1055. [\href{http://dx.doi.org/10.1051/0004-6361:20011123}{CrossRef}]
%20
\bibitem[Miller Bertolami, 2016]{Bertolami16}Miller Bertolami, M.M. New models for the evolution of post-asymptotic giant branch stars and central stars of planetary nebulae.  \textit{Astron. Astrophys.} \textbf{2016},  \emph{588}, A25. [\href{http://dx.doi.org/10.1051/0004-6361/201526577}{CrossRef}]


%21
\bibitem[Hall et al., 2013]{Hall13} Hall, P.D.; Tout, C.A.; Izzard, R.G.; Keller, D. Planetary nebulae after common-envelope phases initiated by low-mass red giants. \emph{Mon. Not. R. Astron. Soc.} \textbf{2013}, \emph{435}, 2048--2059. [\href{http://dx.doi.org/10.1093/mnras/stt1422}{CrossRef}]
%\newpage

%22
\bibitem{Dorman1993} Dorman, B.; Rood, R.T.; O'Connell, R.W. Ultraviolet Radiation from Evolved Stellar Populations. \emph{{arXiv}}\textbf{ 1993}, arXiv: astro-ph/9311022.


%23


\bibitem{Hen1967} Henize, K.G. Observations of Southern Planetary Nebulae. \emph{Astrophys. J. Suppl. Ser.} \textbf{1967},  \emph{14}, 125. [\href{http://dx.doi.org/10.1086/190151}{CrossRef}]









%\bibitem[Renedo et al., 2010]{Renedo10} Renedo I., Althaus L.~G., Miller Bertolami M.~M., Romero A.~D., Corsico A.~H., Rohrmann R.~D., et al. \textbf{2010}, \textit{Astrophys. J.}, 717, 183












\end{thebibliography}
\end{document}